\documentclass[conference]{IEEEtran}
\IEEEoverridecommandlockouts
\usepackage{cite}
\usepackage{amsmath,amssymb,amsfonts}
\usepackage{algorithmic}
\usepackage{graphicx}
\usepackage{textcomp}
\usepackage{csquotes}
\usepackage{subcaption}
\usepackage[ruled,vlined,linesnumbered,resetcount]{algorithm2e}
\usepackage{xcolor}
\def\BibTeX{{\rm B\kern-.05em{\sc i\kern-.025em b}\kern-.08em
		T\kern-.1667em\lower.7ex\hbox{E}\kern-.125emX}}

\begin{document}
	
	\title{Distributed Joint Power and Rate Control for NOMA/OFDMA in 5G and Beyond  \\
		\thanks{This work is supported by the Academy of Finland: (a) ee-IoT n.319009, (b) EnergyNet n.321265/n.328869, and (c) FIREMAN n.326270/CHISTERA-17-BDSI-003; and by JAES Foundation via STREAM project.}
	}

\author{Shiva Kazemi Taskou$ ^* $,  Mehdi Rasti$ ^{*,\dagger} $, Pedro H. J. Nardelli$ ^\dagger $, and Arthur S. de Sena$ ^\dagger $ \\
	$ ^* $Department of Computer Engineering, Amirkabir University of Technology, 
	Tehran, Iran\\
	$ ^\dagger $School of Energy Systems, {Lappeenranta-Lahti University of Technology}, Lappeenranta, Finland \\
	Email: {\{shiva.kt,rasti\}@aut.ac.ir}, pedro.nardelli@lut.fi, arthurssena@ieee.org}

	\maketitle
	\begin{abstract}
		In this paper, we study the problem of minimizing the uplink aggregate transmit power subject to the users' minimum data rate and peak power constraint on each sub-channel for multi-cell wireless networks. To address this problem, a distributed sub-optimal  joint power and rate control algorithm called JPRC is proposed, which is applicable to both non-orthogonal frequency-division multiple access (NOMA) and orthogonal frequency-division multiple access (OFDMA) schemes. Employing JPRC, each user updates its transmit power using only local information. 
		Simulation results illustrate that the JPRC  algorithm can reach a performance close to that obtained by the optimal solution via exhaustive search, with the NOMA scheme achieving a  59\%  improvement on the aggregate transmit power over the OFDMA counterpart. It is also shown that the JPRC  algorithm can outperform existing distributed power control algorithms.  
	\end{abstract}
	
\begin{IEEEkeywords}
		Aggregate transmit power, beyond 5G, OFDMA, NOMA, distributed power control algorithm.
\end{IEEEkeywords}

\section{Introduction}

Orthogonal multiple access (OMA) techniques such as orthogonal frequency-division multiple access (OFDMA) have been the standard employed in previous generation networks.
 In OFDMA, the frequency spectrum is divided into multiple orthogonal sub-channels, each of which is allocated to at most one user at each cell. One of the advantages of OFDMA is that the intra-cell interference can be effectively eliminated \cite{survey-one}. However, performing an orthogonal allocation of  sub-channels results in inefficient usage of frequency resources. On the other hand, non-orthogonal frequency-division multiple access (NOMA) allows multiple users within a cell to operate using the same frequency spectrum. Such a feature makes NOMA a promising technique to improve the frequency efficiency and fulfill the requirements of beyond-5G systems, such as low latency and massive connectivity.
  In NOMA, the signals of different users are decoded sequentially following a given order through successive interference cancellation (SIC). Specifically, each user's symbol is decoded by treating only higher-order users' messages as interference, whereas the interference from lower-order users is removed by SIC   \cite{6G-servey}. Compared to the OFDMA scheme, in NOMA, intra-cell and inter-cell interference management is an important issue. 

Power control is an efficient solution to assist uplink interference management in both NOMA and OFDMA schemes.  Moreover, since mobile and internet of things (IoT) users are battery-powered, the employment of power control strategies has been essential to minimize transmit power in wireless communication networks. To improve energy efficiency, power control solutions minimize the users’ transmit power so that its minimum data rate is met at the lowest possible energy costs.

The existing power control methods for interference management can be classified into (1) power control for NOMA cellular networks \cite{lwc-2018}-\cite{twc-2017} and (2) power control for OFDMA cellular networks  \cite{fwf-2017}-\cite{tgcn-2019}.
Specifically, in \cite{lwc-2018}--\cite{jiot-2018-noma}, power control methods for NOMA single-cell networks are proposed. The authors in \cite{lwc-2018} and \cite{tvt-2018} have proposed the power control methods to maximize the aggregate data rate. In \cite{jiot-2020}, a power control method for maximizing the aggregate data rate in the cognitive IoT is presented.  Authors in \cite{tsp-2020} have proposed a power control algorithm to maximize the weighted sum rate for a single-cell network.
A centralized power control method is proposed in \cite{jiot-2018-noma} to minimize the aggregate transmit power in the uplink. Furthermore, the power control for the downlink of NOMA multi-cell networks is proposed in \cite{lcomm-2018}--\cite{twc-2017}. Authors in \cite{lcomm-2018}--\cite{access-2020} have also proposed centralized power control methods to minimize the aggregate transmit power, whereas in \cite{twc-2017} a distributed power control algorithm is proposed to minimize the aggregate transmit power.

Furthermore, in \cite{fwf-2017} for an OFDMA single-cell network, optimal power control algorithms are proposed to solve a class of aggregate transmit power minimization problems via the water-filling method. 
Also, the authors in \cite{scell-muser} have investigated aggregate transmit power minimization and  aggregate data rate maximization problems for a single-cell network.
The authors in \cite{wf_2013} have investigated the problem of maximizing the aggregate rate subject to power constraint for a single-cell network. To address the formulated problem, the water-filling method was employed.
In \cite{twc-2018}, a power control method has been proposed to maximize the aggregate data rate in single-cell networks.
In \cite{twc-cran-2018}--\cite{jiot-2018}, centralized power control algorithms have been proposed for minimizing the aggregate transmit power of OFDMA multi-cell networks. 

In most existing works \cite{lwc-2018}--\cite{access-2020} and \cite{fwf-2017}--\cite{jiot-2018}, centralized power control algorithms have been proposed for interference management in multi-cell networks. However, distributed approaches, due to requiring users' local information and lower signaling overhead, are preferred to centralized ones. In this regard, in \cite{twc-2017}, a distributed power control algorithm was proposed to minimize the aggregate transmit power in the downlink of a NOMA multi-cell wireless network. Furthermore, in  \cite{tgcn-2019}, distributed power control algorithms were proposed for OFDMA wireless networks. The authors in \cite{tgcn-2019} have proposed a distributed power control method to minimize the aggregate transmit power and offload energy consumption. However, the system model in \cite{tgcn-2019} considered a multi-cell network, where the inter-cell interference was defined as a constant threshold.
Distributed power control algorithms for code-division multiple access (CDMA) cellular networks have also been proposed.
For instance, the target-SINR-tracking power control (TPC) algorithm was proposed in \cite{tpc} to minimize the aggregate transmit power subject to users' minimum data rate.

In this paper, we study the uplink power and rate control problem for both NOMA and OFDMA multi-cell networks. To this end, we formally state the joint power and rate control problem to minimize the aggregate transmit power subject to the users' required data rates and peak power constraints on each sub-channel. To address this problem, we propose a distributed sub-optimal algorithm. 
 
 Specifically,  our main contributions are summarized as:
 \begin{itemize} 
 	\item  Inspired by distributed power control methods for CDMA networks and water-filling method for single-cell OFDMA networks, we propose a distributed joint power and rate control algorithm to address the problem of minimizing the aggregate transmit power subject to users' minimum data rate and peak power constraint on each sub-channel (so-called {JPRC}). The JPRC  algorithm is applicable to both NOMA and OFDMA schemes.
 	
 	\item  Simulation results illustrate that the JPRC  algorithm achieves a performance near to that of the optimal solution obtained by the exhaustive search.  The simulation results also show that our  JPRC  algorithm outperforms the existing centralized and distributed power control algorithms.  Moreover, it is demonstrated that aggregate transmit power for the NOMA  scheme obtains an improvement of  59\% if compared with the OFDMA  scheme. 
 \end{itemize}
 
 The rest of this paper is organized as follows. In Section \ref{system-model}, the system model is proposed.  The optimization problem is formally stated in Section \ref{propblem-statement}. Section \ref{algorithms} introduces our proposed distributed joint power and rate control algorithm.   Finally, simulation results and conclusion are presented in  Section \ref{simulation-results} and Section \ref{conclusion}, respectively.

\section{System Model}\label{system-model}
We consider the uplink of a multi-cell   network consisting of $B$ base stations (BSs)  and $U$ users  denoted  by $\mathcal{B} = \{1,2, \cdots, B\}$ and $\mathcal{U} = \{1,\ldots,U\}$, respectively.  The available  bandwidth is divided into $ C $ sub-channels, represented by the set  $\mathcal{C} = \{1,2, \cdots, C\}$.   The BS   serving user $i$ is denoted by $b_i$ and the set of users served by the  BS $m\in\mathcal{B}$ is denoted by  $\mathcal{U}_{m}$. 
Let $p_i^k$ denote the transmit power of user $i \in \mathcal{U}$  on sub-channel $k \in \mathcal{C}$, the matrix $\boldsymbol{P}=[p_i^k]_{U\times C}$ stands for power allocation of all users over $C$ sub-channels.  
$h^k_{m,i}$ and $N_{m}$ are the path-gain between  user $i$ and BS $m$ and noise power   at BS $m$, respectively. 

The set of interfering users in NOMA and OFDMA schemes is obtained as follows.
\begin{itemize}
\item \textbf{NOMA scheme}:  In the NOMA scheme, we  consider that at each BS $ m $, users with better path-gains are decoded first. Hence, using the SIC technique on each sub-channel $ k $, each user $ i $ assigned to BS $ m $ can successfully remove the interference from users in the same cell whose path-gain is less than $ h ^ k_ {m, i} $, i.e., the interference of all users $  j \in\mathcal{U}_m,~~ \text{if}~ h ^ k_ {m, j} < h ^ k_ {m, i} $ can be canceled successfully \cite{noma}. Therefore, in each sub-channel $ k $,  user $ i $ experiences intra-cell interference only from users whose path-gain is greater than user $ i $'s and inter-cell interference from other cells' users occupying the same sub-channel \cite{noma}. 
We define $ \mathcal{Q}_{m,i}^k $ as the set of users who cause interference on user $ i $ i.e., $ \mathcal{Q}_{m,i}^k=\{\forall j \in\mathcal{U}_m \mid h ^ k_ {m, j} > h ^ k_ {m, i}, ~a_j^k=1 \}\cup \{\forall j \notin\mathcal{U}_m \mid ~a_j^k=1 \} $ in which $ a_j^k $ denotes the allocation of sub-channel $ k $ to user $ j $. If $ a_j^k=1 $, sub-channel $ k $ is allocated to user $j $; otherwise,  $ a_j^k=0 $. 

\item \textbf{OFDMA scheme}: In the OFDMA scheme, each user $ i $ experiences interference from users within other cells that transmit on sub-channel $ k $. So, the set of interfering users for user $ i $ is defined as $ \mathcal{Q}_{m,i}^k=\{\forall j \notin\mathcal{U}_{m} \mid ~a_j^k=1 \} $.

\end{itemize} 

Let $ \zeta_i^k(\boldsymbol{P}) $ be the effective interference of user $ i $ on sub-channel $ k $ which is  obtained by
\vspace{-0.35em}
\begin{equation}\label{effective interference-multi}
\begin{aligned}
\zeta_i^k(\boldsymbol{P})=\dfrac{\sum\limits_{j \in  \mathcal{Q}_{b_i,i}^k}p_j^kh^{k}_{b_i,j}+N_{b_i}}{{h}^k_{b_i,i}},~~\forall i\in \mathcal{U},\forall k\in \mathcal{C},
\end{aligned}
\end{equation} 
where $ \sum\limits_{j \in  \mathcal{Q}_{b_i,i}^k}p_j^kh^{k}_{b_i,j} $ is the  interference on user $ i $.  

The received SINR at BS $b_i$ due to the transmission of  user $i$ on  sub-channel $k$  is given by
\vspace{-0.35em}
\begin{equation}\label{sinr}
\begin{aligned}
\gamma_i^k(\boldsymbol{P}) = \frac{ p_i^k}{\zeta_i^k(\boldsymbol{P})},~~\forall i\in \mathcal{U},\forall k\in \mathcal{C}.
\end{aligned}
\end{equation} 
It is worth noting that the difference between SINR of user $ i $ in NOMA and SINR of user $ i $ in OFDMA holds in the set of interfering users on user $ i $. 

In both NOMA and OFDMA schemes, the achieved data rate of user $i$ on  sub-channel $k$  is expressed as 
\vspace{-0.35em}
\begin{equation}\label{rate}
R_i^k(\boldsymbol{P}) = \log_2\left(1+\dfrac{p^k_i}{\zeta_i^k}\right),~~~\forall i\in \mathcal{U},\forall k\in \mathcal{C}.
\end{equation}
The total data rate of user $ i $ on all allocated sub-channels is obtained by $R_i= \sum\limits_{k \in \mathcal{C}_i}R_i^k $ where  $ \mathcal{C}_i $ is the  allocated sub-channels set to user $ i $, i.e., $ \mathcal{C}_{i}=\{\forall k \in\mathcal{C} \mid ~a_i^k=1 \} $.  
 
  
\section{Problem Statement}\label{propblem-statement}
Given a sub-channel allocation, the power and rate control problem to minimize the aggregate transmit power is formally stated  as 
\begin{equation*}
\begin{aligned}
&\displaystyle \min_{\substack{\boldsymbol{P}}}
&& \sum\limits_{i \in \mathcal{U}}\sum\limits_{k \in \mathcal{C}_i}p_i^k\\
&\text{s.t.}
&&\mathrm{C1}:~~\sum\limits_{k \in \mathcal{C}_i}R_i^k \ge R_{i}^{\mathrm{min}},~~~\forall i \in \mathcal{U},\\
\end{aligned}
\end{equation*} 
\begin{equation}\label{peak-power}
\begin{aligned}
&
&&\mathrm{C2}:~~0\leq p_i^k \leq \bar{p}_i^k,~~~~~~~~~~~\forall i \in \mathcal{U},\forall k \in \mathcal{C}_i,
\end{aligned}
\end{equation}                       
where constraint $\mathrm{C1}$  implies that the minimum data rate denoted by  $ R_{i}^{\mathrm{min}} $ should be met by each user $i$. Constraint  $\mathrm{C2}$ corresponds to the constraint of peak power on each  sub-channel i.e., $\bar{p}_{i}^{k}$. Also, $\mathrm{C2}$  implies that the transmit power on each sub-channel should be a non-negative value.

Problem  \eqref{peak-power} is defined in both NOMA and OFDMA schemes. 
The research works \cite{jiot-2018-noma}--\cite{twc-2017} investigated the aggregate transmit power minimization problem for NOMA scheme, while \cite{fwf-2017}--\cite{scell-muser},  \cite{twc-cran-2018}--\cite{jiot-2018}, and \cite{tgcn-2019} studied this problem for OFDMA scheme.
Problem \eqref{peak-power} is non-convex, thus there is no polynomial-time algorithm to solve it optimally. In  what follows,  we propose a sub-optimal algorithm to address this problem in a distributed manner.

\section{Proposed Distributed Joint Power and Rate Control Algorithm}\label{algorithms}
To address problem \eqref{peak-power} in a distributed manner, there are some notable facts as follows.
\begin{enumerate}
	\item In multi-cell networks,   the existence of interference in SINR of user $ i $ i.e.,  \eqref{sinr} makes the data rate function in \eqref{rate} non-convex. Thus, due to the non-convexity of constraint $ \mathrm{C1} $, problem \eqref{peak-power} is non-convex.  The optimal solution of problem \eqref{peak-power} for multi-cell networks can be obtained via exhaustive search, which is not practical due to the high computational complexity.
	
	\item For single-cell NOMA networks, due to the sharing of sub-channels at each cell, the set of interfering users is not empty, so problem \eqref{peak-power} is still non-convex. However, for single-cell OFDMA networks, since the set of interfering users is empty, the data rate function in \eqref{rate} becomes convex, which makes problem \eqref{peak-power} convex. 
	
	\textbf{Lemma 1.}
	Given a single-cell OFDMA network, the optimal solution of problem \eqref{peak-power} is obtained by the water-filling method proposed in \cite{fwf-2017}. Employing the water-filling method, for each user $ i $, there are optimal values of $ k^*_i $ and  $ S_i $ denoting the index of the last sub-channel with positive power and the last sub-channel on which user $ i $ transmits with peak power.  Having $ k^*_i $ and $ S_i $, 	the optimal transmit power of user $ i $ on each sub-channel $ k $   is obtained  by 
	\begin{equation}\label{water-filling}
	\begin{aligned}
	p_i^k=\begin{cases}
	\bar{p}_i^k, ~~~~~~~~~~~~~~~~~~~~1\leq k\leq S_i\\
	\left({2^{R_i^{\mathrm{min}}}}{\prod_{{k=1}}^{k^*}\dfrac{N_{b_i}}{h^k_{b_i,i}}}\right)^{\dfrac{1}{k^*_i}}\!-\!\dfrac{N_{b_i}}{h^k_{b_i,i}},~ S_i<k\leq k^*_i,
	\end{cases}
	\end{aligned}
	\end{equation}
	where it is assumed that the sequence of $ \{{h^k_{b_i,i}}/{N_{b_i}}\}_{k=1}^{\mid \mathcal{C}_i\mid} $ is sorted 	monotonically decreasing, i.e., $ {h^1_{b_i,i}}/{N_{b_i}}>{h^2_{b_i,i}}/{N_{b_i}}>\cdots>{h^k_{b_i,i}}/{N_{b_i}} $.
	
	\textit{Proof.}
	Lemma 1 is proved in \cite{fwf-2017}.
	
	\item To address problem \eqref{peak-power} in multi-cell networks, the sub-optimal power control algorithms proposed in \cite{jiot-2018-noma}--\cite{twc-2017} and \cite{twc-cran-2018}--\cite{jiot-2018} are centralized approaches in which the global information is needed for calculating users' appropriate transmit power. In current wireless networks and IoT, the centralized approaches are not practical because of the existence of a huge number of users. Thus,  distributed algorithms to address problem  \eqref{peak-power}  in which each user obtains its transmit power knowing local information is interesting.
	Inspired by the water-filling method for single-cell OFDMA networks and distributed power control methods for CDMA networks, we propose the distributed JPRC  algorithm to address problem \eqref{peak-power} for both NOMA and OFDMA schemes in multi-cell networks.
	
\end{enumerate}

Let $ k^*_i $ and $ S_i $ denote the index of last sub-channels with positive power and the last index of sub-channel on which user transmits with peak power $ \bar{p}_i^k $, respectively. To address problem \eqref{peak-power} for multi-cell networks, obtaining $ k^*_i $ and $ \mathcal{S}_i $ in a distributed manner is challenging. To obtain $ k^*_i $ and $ \mathcal{S}_i $, we propose JPRC algorithm which is explained next.

In problem \eqref{peak-power}, constraint $ \mathrm{C1} $ guarantees a minimum data rate, $ R_i^{\mathrm{min}} $ for each user $ i $. In contrast to the CDMA networks, there is no one-to-one relation between $ R_i^{\mathrm{min}} $ and SINR on each sub-channel $ k $   in NOMA and OFDMA networks.
To address problem \eqref{peak-power} in a distributed manner, first, we divide  $ R_i^{\mathrm{min}} $ between allocated sub-channels of user $ i $, $ \mathcal{C}_i $. By doing so, the target data rate of user $ i $ on each sub-channel $ k\in\mathcal{C}_i $ denoted by $ \widehat{R}_i^k $  is obtained. 
Inspired by water-filling method for single-cell OFDMA networks, we calculate $ \widehat{R}_i^k $ as follows.

Substituting the transmit power obtained by \eqref{water-filling} in data rate function \eqref{rate}, we have $ R_i^k=\log_2\Big({{({2^{R_i^{\mathrm{min}}}}{\prod_{k=1}^{k^*_i}\zeta_i^n})^{{1}/{k^*_i}}}}/{\zeta_i^k}\Big) $, where in the initial step $ k^*_i $ is set to $ k^*_i=\mid\mathcal{C}_i\mid $. Employing some mathematics, the data rate for user $ i $ on sub-channel $ k $ is obtained by
\vspace{-0.75em}
\begin{equation}\label{target-rate}
\begin{aligned}
R_i^k\!=\!\log_2\left(\dfrac{1}{\zeta_i^k}\right)+\dfrac{1}{k^*_i}\left(R_i^{{\mathrm{min}}}-\sum_{n=S_i+1}^{k^*_i}\log_2\left(\dfrac{1}{\zeta_i^n}\right)\right).
\end{aligned}
\end{equation}
Using \eqref{target-rate}, when the allocated sub-channel to user $ i $ is good (i.e., low effective interference is caused to the user), it can achieve a higher data rate; therefore, the data rate on that sub-channel increases and vise versa. 

By employing \eqref{target-rate}, the data  rate on some sub-channels may be a negative value; to avoid this,  data   rate on that sub-channel $k$ is set to zero (i.e., $R_i^k=0, ~\text{if}~ R_i^k<0$), then the value of $ k^*_i $ decrements by one (i.e., $k^*_i=k^*_i-1$).

Furthermore, based on the peak power on each sub-channel $k$, the maximum achievable data rate of user $ i $ on each sub-channel $ k $ is expressed as
\vspace{-0.75em}
\begin{equation}\label{rate-peak}
\begin{aligned}
\bar{R}_i^k=\log_2\left(1+\dfrac{\bar{p}_i^k}{\zeta_i^k}\right).
\end{aligned}
\end{equation}
To guarantee peak power constraint $ \mathrm{C2}  $ in problem \eqref{peak-power}, if on sub-channel $ k $, $ {R}_i^k>\bar{R}_i^k $, then $ {R}_i^k $ is set to $ {R}_i^k=\bar{R}_i^k $ and the value of $ S_i $ initially setting to $ S_i=0 $ increments by one, i.e., $ S_i=S_i+1 $.

According to \eqref{target-rate} and \eqref{rate-peak},  the target data rate for user $i\in \mathcal{U}$ on each sub-channel $ k $ is calculated by
\begin{equation}\label{target-rate-peak}
\begin{aligned}
\widehat{R}_i^k=\begin{cases}
\bar{R}_i^k,~~~~~~1 \leq k\leq S_i,\\
R_i^k,~~~~~~S_i<k\leq k^*_i,
\end{cases}.
\end{aligned}
\end{equation}

These steps recur iteratively  until the  data rate on each sub-channel $k=\{1,2,\cdots,k_i^*\}$ be a non-negative value and the total  data rate for user $i$ on its allocated sub-channel equals to minimum data rate i.e., $ \sum_{k=1}^{k^*_i} \widehat{R}_i^k=R_i^{\mathrm{min}}$. 

Then, the value of target data rate for user $ i $ on each  sub-channel $k$ is mapped to target SINR on that sub-channel by
\begin{equation}\label{target sinr updating}
\begin{aligned}
\widehat{\gamma}_i ^k =2^{\widehat{R}_i ^k}-1,~~~~\forall i\in\mathcal{U},~~~\forall k=1,2,\cdots,k^*_i.
\end{aligned}
\end{equation}

By obtaining the target SINR through \eqref{target sinr updating}, the optimal  transmit power of user $ i $ on sub-channel  $k$ can be updated   according to the TPC algorithm in a distributed manner as 
\begin{equation}\label{power updating}
\begin{aligned}
p_i^k =  \begin{cases}
\bar{p}_i^k,~~~~~~~~~~~~~1\leq  k\leq S_i, \\
\widehat{\gamma} _i^k {\zeta_i^k}, ~~~~~~~~~~ S_i<k\leq k^*_i.
\end{cases}.
\end{aligned}
\end{equation} 
Our proposed {JPRC} is summarized in Algorithm 1.

\begin{algorithm}\label{JPRC}
	\SetKwFunction{Range}{range}
	\SetKw{KwTo}{in}\SetKwFor{For}{for}{\string:}{}%
	\SetKwIF{If}{ElseIf}{Else}{if}{:}{elif}{else:}{}%
	\SetAlgoNoEnd
	\SetAlgoNoLine%
	\textbf{Input:}
$p_i^k(1)=0 , \widehat{\gamma}_i^k(1)=2^{\frac{R^{\mathrm{min}}_i}{\mid \mathcal{C}_i \mid}}-1$, $\forall i\in\mathcal{U}$, $\forall k\in\mathcal{C}_i$, $ {S}_i(1)=0 $, $ k^*_i(1)=\mid \mathcal{C}_i \mid $, $ \mathrm{MinRate_i(1)=R_i^{\mathrm{min}}} $, t$ =1 $\\

\textbf{repeat}\\
\Indp
	 \For{\emph{each} $i \in\mathcal{U} $ }{ \While{$ 
				\sum_{k=1}^{k^*_i} \widehat{R}_i^k (t)<R_i^{\mathrm{min}}$}{ \textbf{repeat}\\
					\Indp
					 {\For{\emph{each} $k =S_i(t)+1,\cdots,k^*_i(t)  $}
					{ Obtain $\zeta_{i}^k (t)$  by \eqref{effective interference-multi}.\\
					 Sort the sequence of $\{\dfrac{1}{\zeta_{i}^k (t)}\}_{k=1}^{k^*_i(t)}$  in a descending order. \\
					 Obtain $ R_i^k (t) $  by \eqref{target-rate}.\\

				}}	 \textbf{end of for.} \\
				 \If{$ R_i^k (t)<0 $}{$ {R}_i^k (t)=0 ~\text{and}~k^*_i(t)=k^*_i(t)-1$.}
				 \textbf{end of if.}\\
			\Indm
				 \textbf{until $ R_i^k(t)\geq 0,~\forall k =S_i(t)+1,\cdots,k^*_i(t) $.}\\
			
				 \For{\emph{each} $ k =S_i(t)+1,S_i(t)+2,\cdots,k^*_i(t) $}
				 {Obtain  maximum data rate ($ \bar{R}_i^k  (t)$)  by \eqref{rate-peak}.\\
					 Obtain target  data rate ($ \widehat{R}_i^k (t) $)  by \eqref{target-rate-peak}.\\
					 \If{$   \widehat{R}_i^k  (t) \geq \bar{R}_i^k  (t)$}{ Update $ \mathrm{MinRate}_i(t+1) $ by $ \mathrm{MinRate}_i(t)-\sum_{k=1}^{S_i(t)}\bar{R}_i^k(t)$ and   $S_i(t)=S_i(t)+1$.}				
					 \textbf{end of if.} 
			 } \textbf{end of for.} \\
		  Set $ k^*_i(t)=\mid \mathcal{C}_i \mid $ and go to step 4.\\
		}	 \textbf{end of while.}
	} \textbf{end of for.}\\	
	\Indp
	\Indm
	\Indm
	\Indp
	 Obtain   $\widehat{\gamma}_i^k (t)$  by \eqref{target sinr updating}.\\
	 Update transmit power,  $p_i^k (t)$  by \eqref{power updating}.\\
	 t $ \leftarrow $ t+1 \\
	\Indm	
	 \textbf{until} convergence
	\caption{Our Proposed JPRC Algorithm }
	
\end{algorithm}
\textbf{Remark.}
	JPRC  algorithm is distributed since each user updates its transmit power by knowing its uplink path-gain and total interference at the BS. Therefore, the users do not need to know other users' path-gains, minimum data rate, and peak-power on each sub-channel.
 
\textbf{Theorem 1.}
	The JPRC  algorithm has at least one fixed-point for both NOMA and OFDMA schemes.

\textit{Proof.}
	According to \emph{Brouwer fixed-point theorem}, if $\mathcal{X}$ be a nonempty, compact, and convex subset of $\mathcal{R}^n$,  every continuous function $f : \mathcal{X} \longrightarrow \mathcal{X}$ mapping $\mathcal{X}$ into itself has a fixed point \cite{fixed-point}. As aforementioned, in JPRC, $p_i^k \in [0,\bar{p}_i^k]$ ($\forall i \in \mathcal{U}, \forall k \in \mathcal{C}_i$) where $[0,\bar{p}_i^k]$ is a nonempty,  compact, and convex set \cite {boyd}.
	Furthermore,   function \eqref{power updating} maps $[0,\bar{p}_i^k]$ into itself. 
		Now, we should prove that the function \eqref{power updating} is a continuous function. To this end, we prove that \eqref{target-rate} is a continuous function.  From \cite{boyd}, we have that $ \log_2(1/\zeta_i) $ is continuous on $[0,\bar{p}_i^k]$. Then, its exponential, i.e., \eqref{target sinr updating}, is a continuous function. Additionally, \eqref{effective interference-multi} is a continuous function as well. As a result,  function \eqref{power updating} is a continuous function. Thus, our proposed JPRC algorithm  has at least one fixed-point, which completes the proof.

 \textbf{Theorem 2.} 	
 	In the single-channel scenario, the JPRC  algorithm obtains the optimal solution for problem \eqref{peak-power}.
  
 \textit{Proof.}
 	In the single-channel scenario in which each user at each cell can occupy a single sub-channel, equation \eqref{target-rate} equals to $ R_i^k={R_i^\mathrm{min}} $. Then, the target SINR on that sub-channel $ k $ is obtained by $ \widehat{\gamma}_i ^k =2^{R_i^\mathrm{min}}-1 $. Therefore, using \eqref{power updating} the transmit power of each user $ i $ on sub-channel $ k $ is given by 
 	\begin{equation}\label{single-channel}
 	\begin{aligned}
 	p_i^k =  \begin{cases}
 	\bar{p}_i^k,~~~~~~~~~~~~~~~~~~~~~~~~~~ k\leq S_i, \\
 	\left(2^{R_i^\mathrm{min}}-1\right) {\zeta_i^k}, ~~~~~~~~~~ S_i<k\leq k^*_i.
 	\end{cases}
 	\end{aligned}
 	\end{equation} 
 	Based on frameworks for power updating functions in \cite{standards}, if a power updating function falls into the standard type-I framework, the power updating function has a unique fixed point to which converges. An power updating  function $ f(\boldsymbol{P}) $ falls into standard type-I if it satisfies the following two monotonicity and scalability conditions
 	\begin{itemize}
 		\item Monotonicity: if $ \boldsymbol{P^{\prime}} \leq \boldsymbol{P} $, then $ f(\boldsymbol{P^{\prime}}) \leq f(\boldsymbol{P}) $.
 		\item Scalability: for all $ \alpha>1 $, $ f(\boldsymbol{\alpha P}) < \alpha f(\boldsymbol{P}) $.
 	\end{itemize}
 	
 	It can be easily proved that our power updating function in \eqref{single-channel} satisfies the monotonicity and scalability conditions, so it is a standard type-I function, which completes the proof.

\begin{figure*}	 
	\begin{subfigure}{0.3\linewidth}
		\includegraphics[width=1\linewidth] {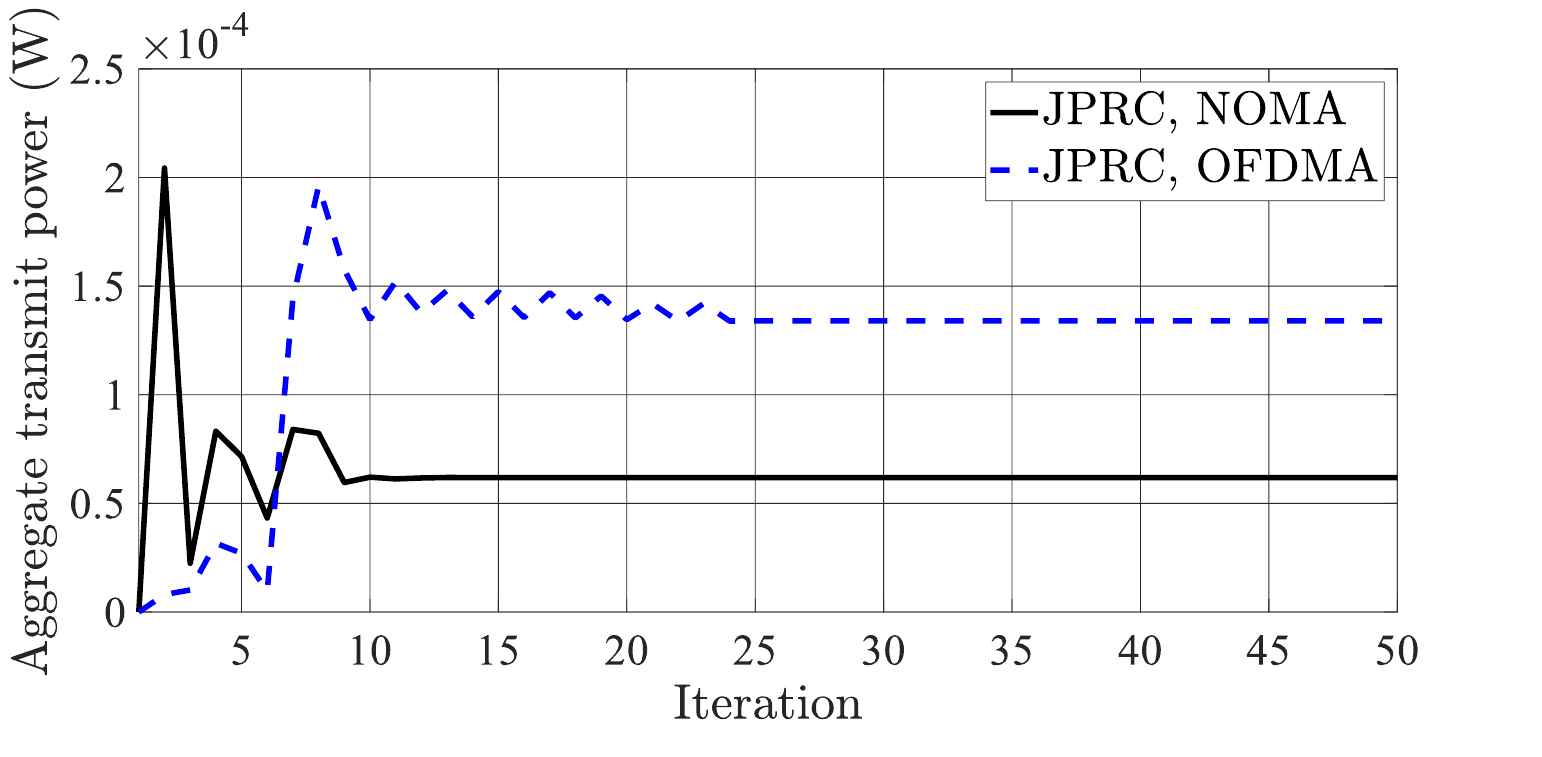}  
		\caption*{Fig. 1: \small Convergence of JPRC  for    NOMA and  OFDMA schemes}
		\label{convergence}
	\end{subfigure}
	\qquad
	\begin{subfigure}{0.3\linewidth}
		\includegraphics[width=1\linewidth]{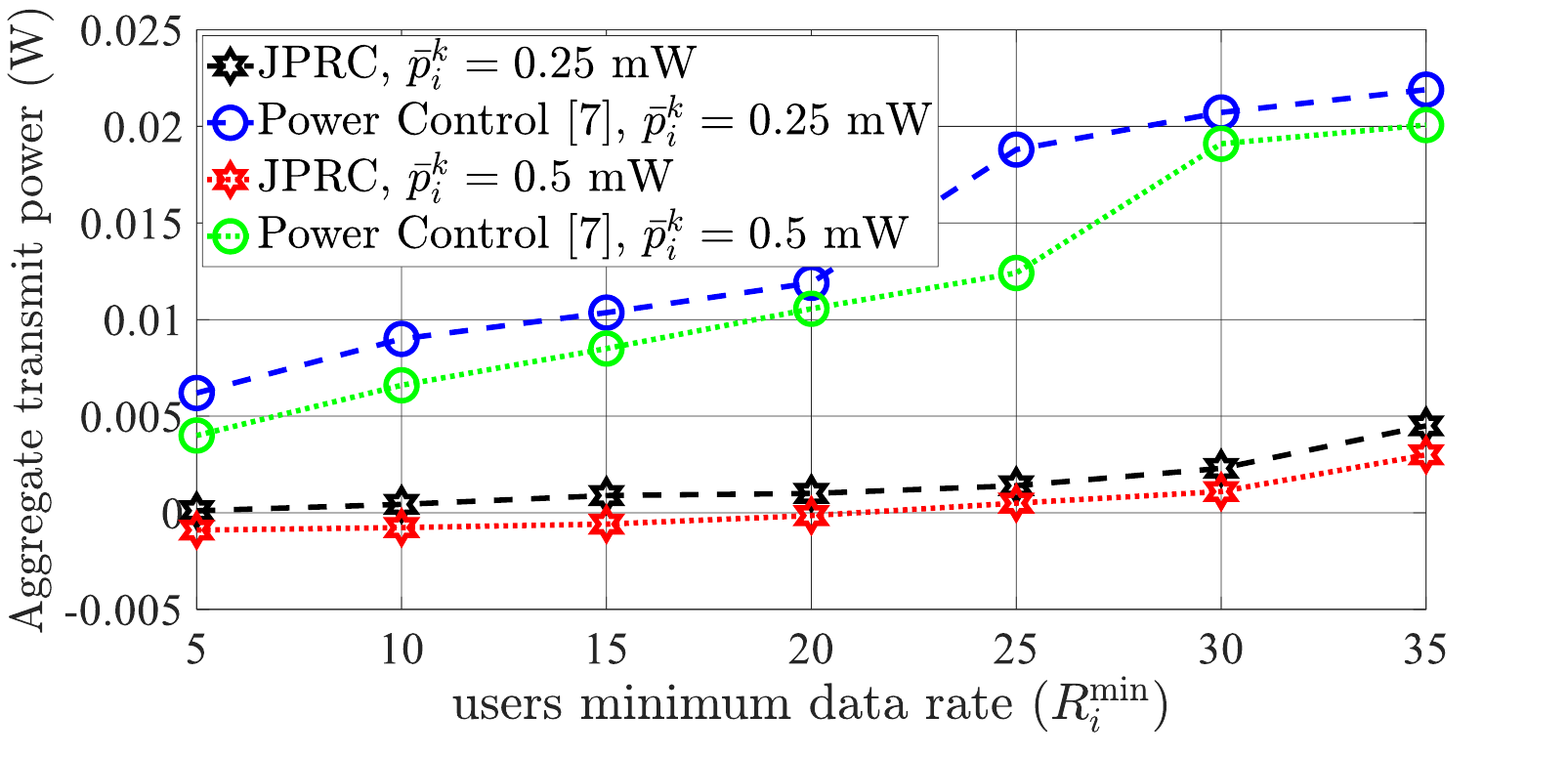}
		\caption*{Fig. 2: \small Comparison of  JPRC  with algorithm proposed in \cite{jiot-2018-noma}  
			for NOMA scheme
		}\label{noma_multi_peak}
	\end{subfigure}
	\qquad 
	\begin{subfigure}{0.3\linewidth}
		\includegraphics[width=1\linewidth] {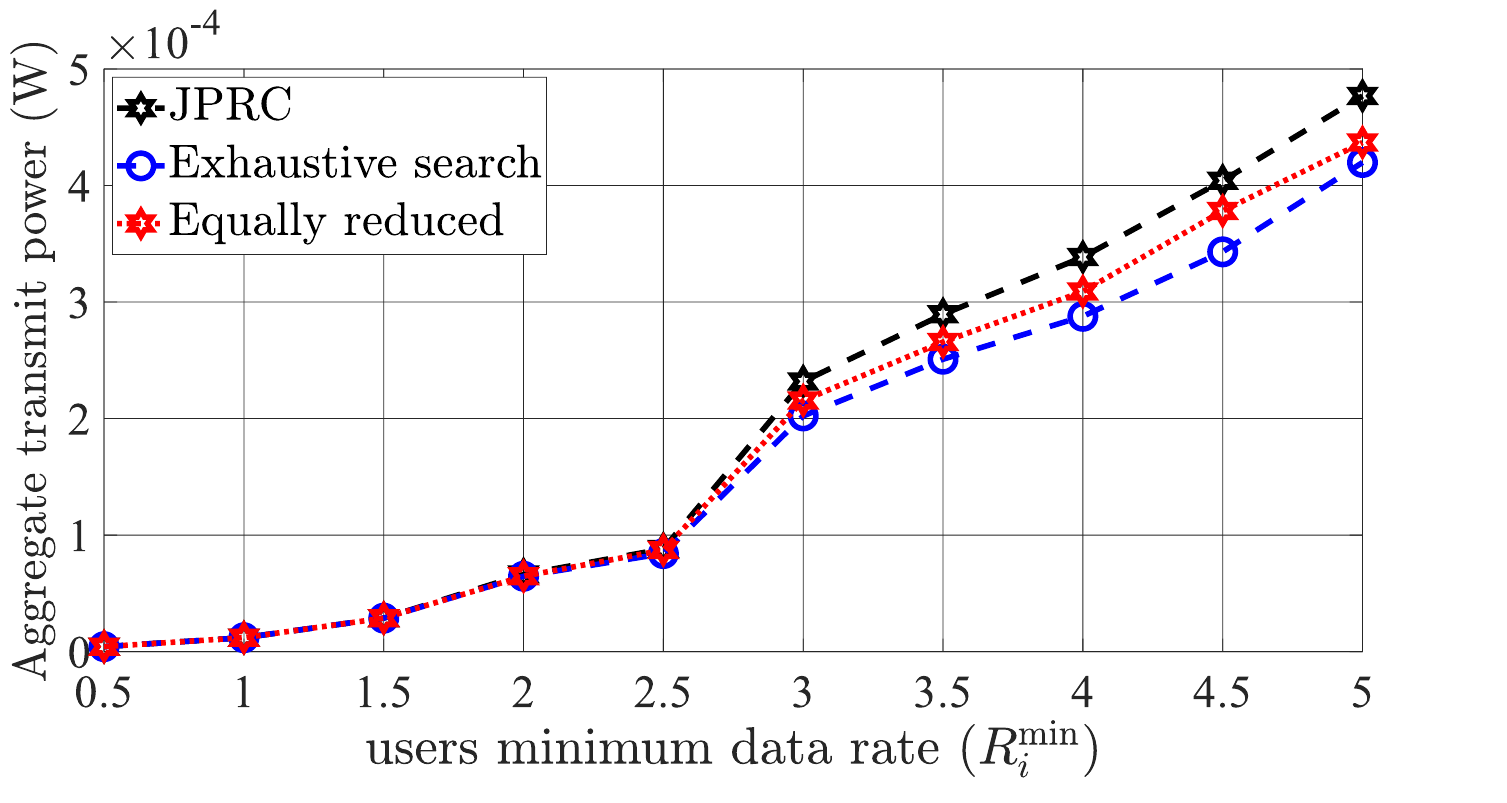}  
		\caption*{Fig. 3: \small Comparison of  {JPRC}  with the exhaustive search and equally reduced power control methods for NOMA scheme}
		\label{multi_power_minimum_rate_peak_optimal}
	\end{subfigure}
	\vspace{-0.8 em}
\end{figure*}
 
\section{Simulation Results}\label{simulation-results}
To evaluate the performance of the JPRC  algorithm, we consider a multi-cell network consisting of $4$ cells with  $500\mathrm{m} \times 500\mathrm{m}$ coverage area in an area of $1000\mathrm{m} \times 1000\mathrm{m}$ which  $4$ users are randomly located in each cell.
Similar to \cite{pathgain}, the uplink path-gain from each user $i$ to each base station $m$ is modeled by $h_{m,i}^{k}=x(k)d_{m,i}^{-\alpha}$, where $d_{m,i}$ is the distance between BS $m$ and user $i$,   $x(k)$ is a random value that is generated by the Rayleigh distribution, and $\alpha = 3 $ is the path loss exponent. All the simulation parameters are stated here unless stated otherwise. The mask transmit power of users set to $ \bar{p}_i^k=0.25\mathrm{mW} $. Users' minimum data rate is $ R_i^{\mathrm{min}}=5\mathrm{bps/Hz} $.   The noise power, $ N_{m} $ is set to $ 10^{-14} \mathrm{W} $.  Additionally, for the OFDMA scheme, a number of $ 100 $ sub-channels are equally allocated to users in each cell. Whereas in the NOMA scheme, the whole sub-channel set is allocated to each user. 

In what follows, first, we show the convergence of JPRC; then, we evaluate the performance of our proposed JPRC  algorithm for  NOMA and OFDMA schemes. Finally, we compare the aggregate transmit power for the NOMA scheme with that of the OFDMA scheme.
We average from 500 independent snapshots with different users' locations and different path-gains to generate each curve.
\subsection{Convergence of JPRC }
Fig. 1  illustrates the convergence of our proposed JPRC  algorithm in terms of the aggregated transmit power versus the number of iterations for NOMA and OFDMA schemes. From Fig. 1, it can be observed that our proposed algorithm converges to one of its fixed-points after about 20  iterations. Besides, it can be seen that due to sharing sub-channels among all users, the aggregate transmit power obtained by the NOMA scheme is less than that of the OFDMA scheme.
\begin{figure*}	 
	\begin{subfigure}{0.3\linewidth}
		\includegraphics[width=1\linewidth]{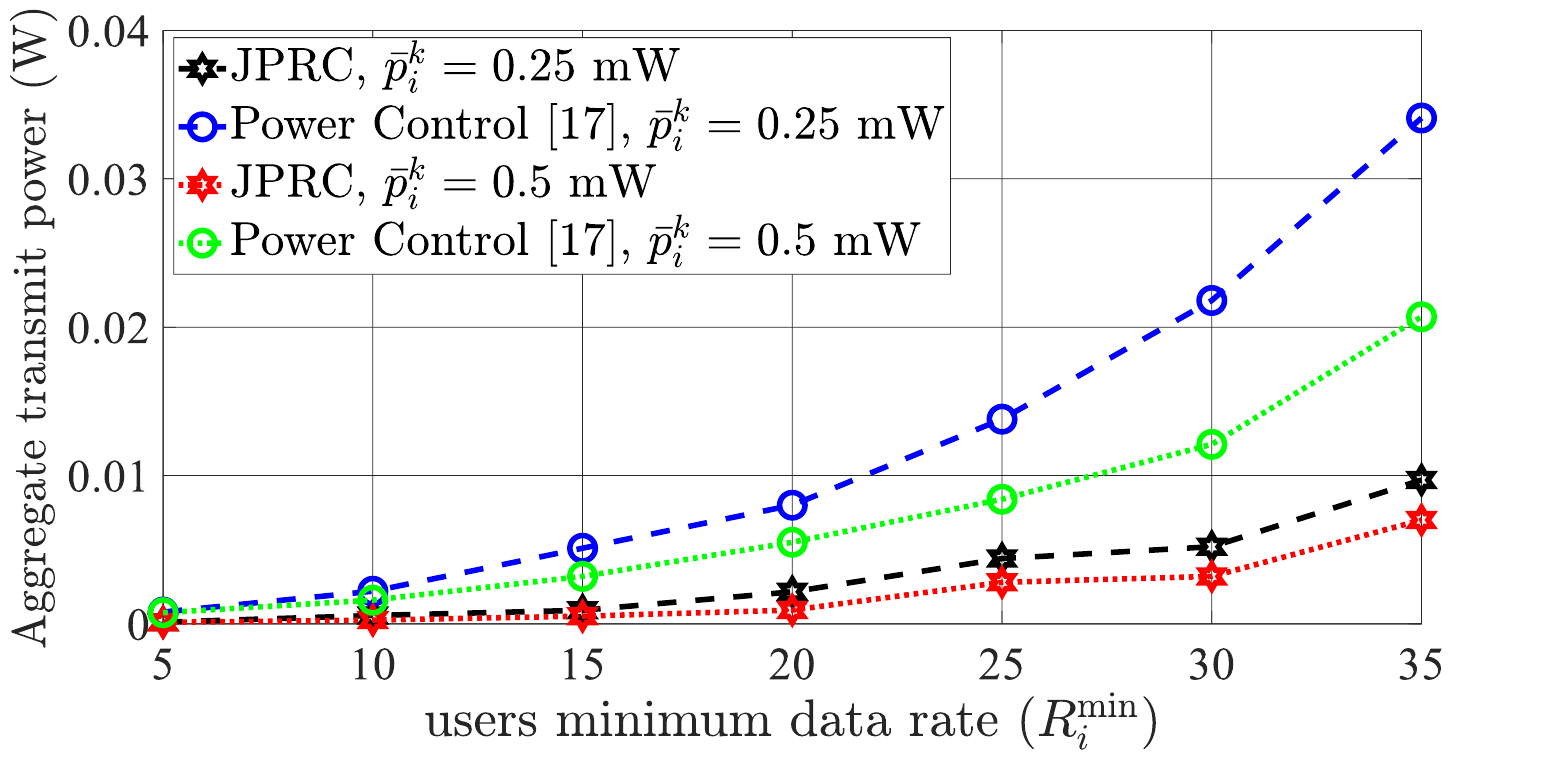}  
		\caption*{Fig. 4: \small Comparison of  {JPRC}  with distributed power control algorithm proposed in \cite{tgcn-2019}  for OFDMA scheme }
		\label{multi_power_minimum_rate_peak}
	\end{subfigure}
	\qquad
	\begin{subfigure}{0.3\linewidth}
		\includegraphics[width=1\linewidth]{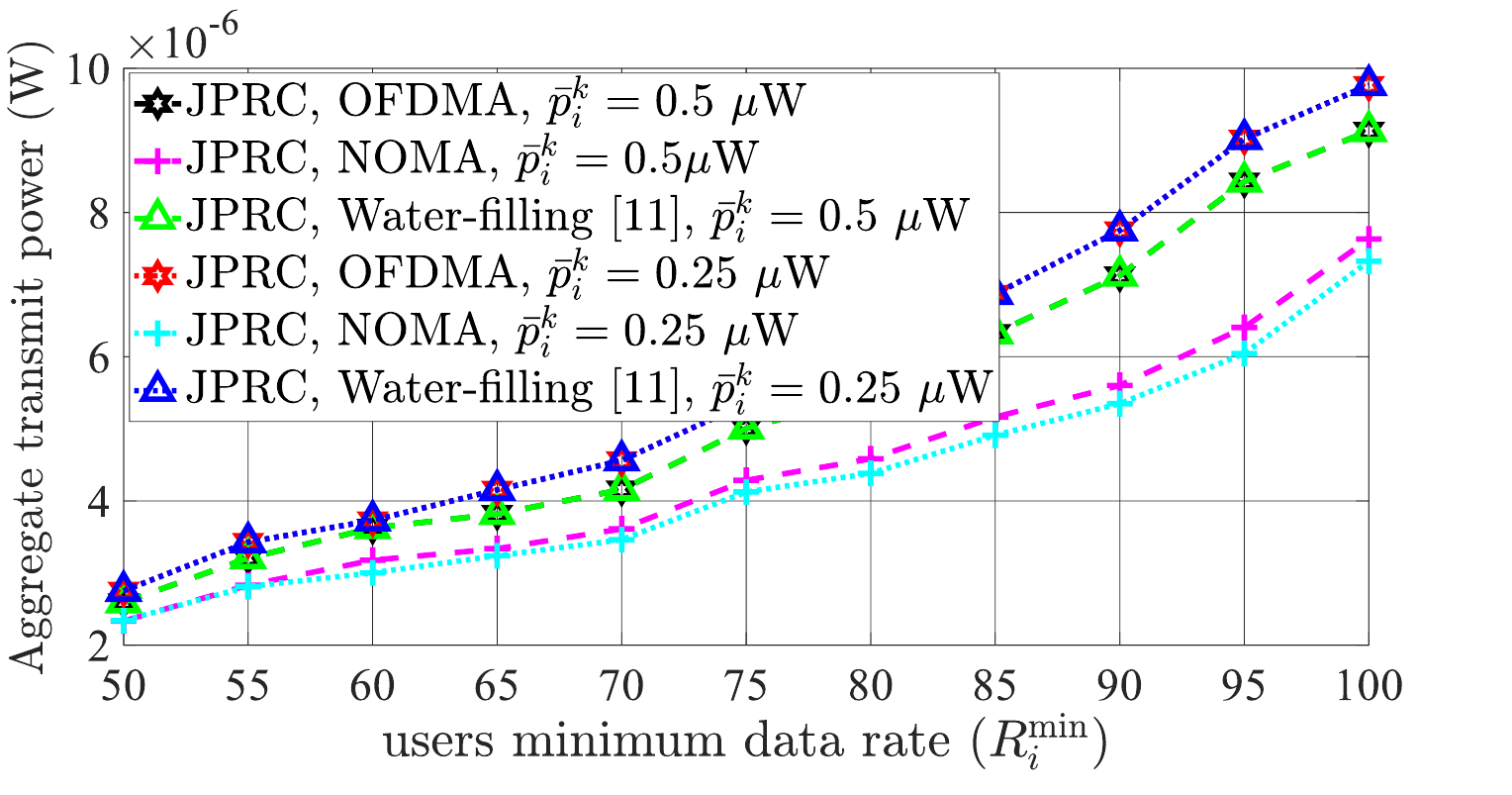}  
		\caption*{Fig. 5: \small Comparison of {JPRC} with the optimal solution in \cite{fwf-2017} for NOMA and OFDMA schemes in a single-cell nework}
		\label{jprc_pp_single_power_minimum_rate}
	\end{subfigure}
	\qquad 
	\begin{subfigure}{0.3\linewidth}
		\includegraphics[width=1\linewidth] {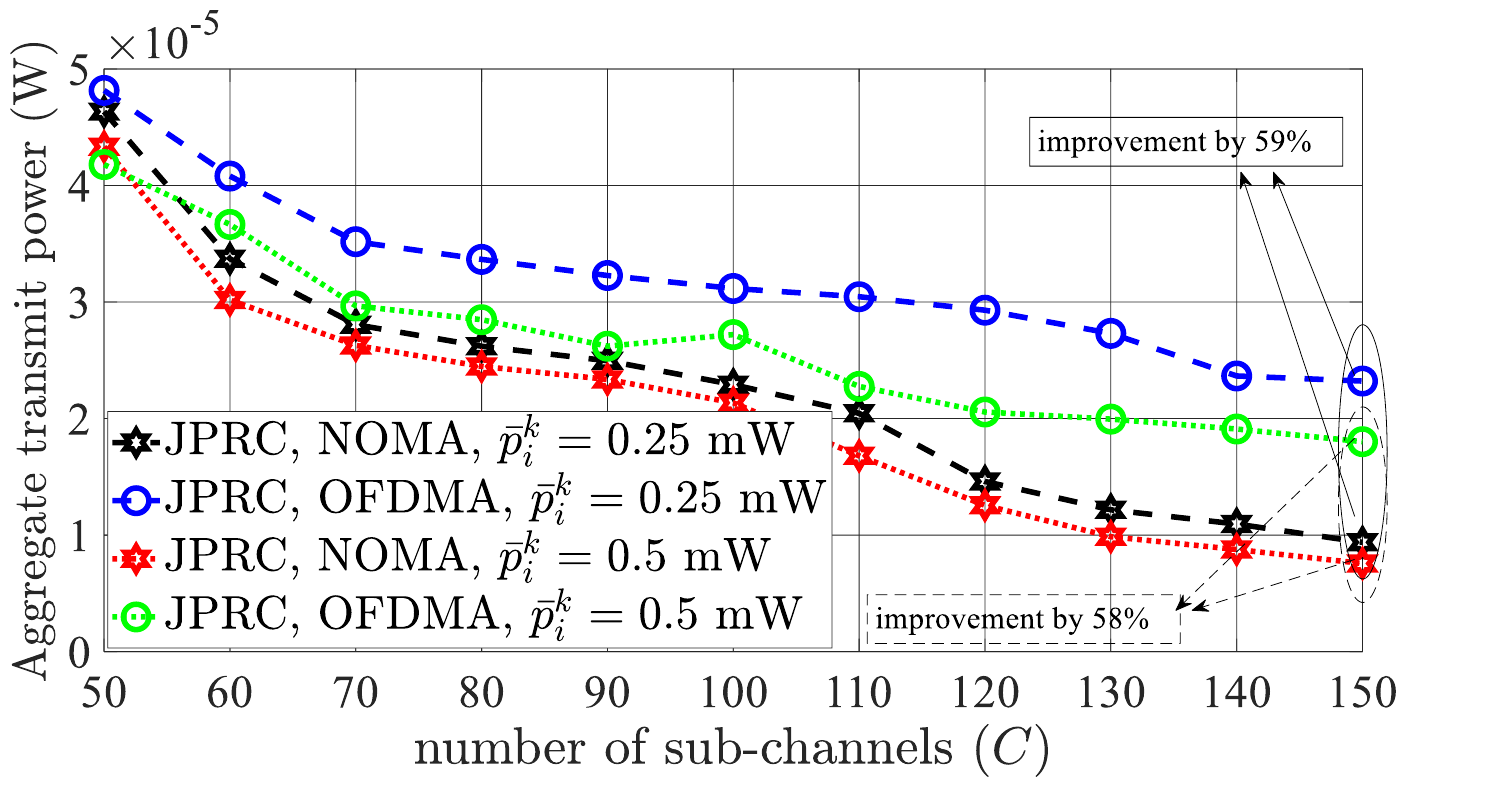}  
		\caption*{Fig. 6: \small Aggregate transmit power of JPRC  for NOMA and OFDMA schemes in a multi-cell network }
		\label{noma_ofdma_single_peak}	
	\end{subfigure}
	\vspace{-0.8 em}
\end{figure*}

\subsection{ Evaluation of  the JPRC  Algorithm for NOMA Scheme}
In this sub-section,  to evaluate the performance of JPRC in NOMA multi-cell networks, we compare it with the centralized power control algorithm proposed in \cite{jiot-2018-noma},  the optimal solution obtained by the exhaustive search, and the equally reduced power control method.

Fig. 2 compares the performance of JPRC  algorithm with power control algorithm proposed in \cite{jiot-2018-noma} when the users' minimum data rate increases from $ 5 \mathrm{bps/Hz} $  to $ 35\mathrm{bps/Hz} $. From Fig. 2, we can observe that when the minimum data rate increases, the aggregate transmit power increases. Likewise, it can be seen in Fig. 2, with increasing the peak power on each sub-channel, the aggregate transmit power is reduced. 
Moreover, the JPRC  algorithm outperforms the centralized power control method in \cite{jiot-2018-noma} since the high-SINR scenario is considered in \cite{jiot-2018-noma} which is not a realistic assumption for beyond 5G.

In Fig. 3, we compare our proposed JPRC  algorithm with the optimal solution obtained by exhaustive search and equally reduced power control methods. For the exhaustive search, we quantize the transmit power on each sub-channel, considering a sufficiently small step size. In the exhaustive search method, all possible quantized power values are searched.
In the equally reduced power control method, each user transmits equal power on its allocated sub-channels. In this method, using the exhaustive search, the users' transmit power on each sub-channel is reduced equally to satisfy their minimum data rate.
To generate Fig. 3, we consider two cells with $ 500\times 500 $ coverage area in which one user is randomly located. Also, to each user,  2  sub-channels are allocated, and the peak power on each sub-channel is set to $ 0.5\mathrm{\mu W} $.  
From Fig. 3, it can be seen that JPRC obtains the same solution as the exhaustive search and the equally reduced power control methods when the users' minimum data rate decreases.

\subsection{ Evaluation of  the JPRC  Algorithm for OFDMA Scheme }
In this sub-section, for the OFDMA scheme,  we compare the performance of the JPRC  algorithm with the distributed power control algorithm proposed in \cite{tgcn-2019}.
Fig. 4 illustrates the performance of our proposed JPRC  algorithm compared to the distributed power control algorithm proposed in \cite{tgcn-2019},  when  the users' minimum data rate  increases from $5\mathrm{bps/Hz}$ to $35\mathrm{bps/Hz}$ by setting the  peak power on each sub-channel $ k $ to $ \bar{p}_i^k=0.25\mathrm{mW} $  and $ \bar{p}_i^k=0.5\mathrm{mW} $.
We can observe from Fig. 4,   due to transmitting with high power on good sub-channels, increasing the peak power results in decreasing the aggregate transmit power.   Moreover, in Fig. 4, with increasing $R^{\mathrm{min}}_i$, since users should transmit with a higher power to reach their minimum data rate, the aggregate transmit power increases.  Furthermore, the aggregate transmit power obtained by JPRC  is lower than that of power control proposed in \cite{tgcn-2019} in  Fig. 4. The reason is that in \cite{tgcn-2019}, to convexify the power control problem, the interference at each sub-channel is set to a constant threshold. While, with power control, the interference in each sub-channel is different and may be lower than a predefined threshold.

\subsection{Comparing Performance of  NOMA  With OFDMA }

In this sub-section, we compare the aggregate transmit power of the JPRC  algorithm for the NOMA scheme with that of the OFDMA scheme.

Considering a single-cell network with a central BS and 4 randomly located users, in Fig. 5, we compare the performance of our proposed {JPRC}   algorithm for NOMA and OFDMA schemes with the optimal solution by the water-filling method proposed in \cite{fwf-2017} for OFDMA scheme.  
In  Fig. 5,  we increase the user's minimum data rate from $ 50\mathrm{bps/Hz} $  to  $ 100\mathrm{bps/Hz} $ setting number of sub-channels to $ 100 $. 
It can be observed from  Fig. 5 with increasing $R^{\mathrm{min}}_i$, since the user has to transmit with higher transmit power to obtain its minimum data rate, the aggregate transmit power increases. 
Furthermore, from Fig. 5, we can observe that increasing the peak power on each sub-channel decreases the aggregated transmit power. Since when the peak power increases, the user can transmit with higher transmit power on good sub-channels. Besides, it can be seen that considering the OFDMA scheme, JPRC obtains the same solution as the water-filling method proposed in \cite{fwf-2017}. Moreover, JPRC for the NOMA scheme obtains a lower aggregate transmit power than the JPRC   for the OFDMA scheme due to sharing the whole spectrum by all users.

Additionally, in Fig. 6, we compare the aggregate transmit power of JPRC in NOMA and OFDMA schemes for a multi-cell network. From Fig. 6, we can observe that since all users can share the whole spectrum in the NOMA scheme, the aggregate transmit power is reduced. Hence, it can be seen that the NOMA scheme improves the aggregate transmit power by   59\%    compared to the OFDMA.

\section{Conclusion}\label{conclusion}
In this paper, we proposed a distributed JPRC  algorithm for NOMA/OFDMA wireless networks to address the aggregate transmit power minimization problems by considering peak power constraint on each sub-channel and minimum data rate for each user.  Employing JPRC, which has at least one fixed-point, each user updates its transmit power, knowing local information. Simulation results illustrated that JPRC  outperforms the existing power control algorithms and obtains a near solution to the optimal solutions. Besides, via simulation results, we showed that the NOMA scheme decreased the aggregate transmit power by 59\%  compared to the OFDMA scheme. 


\vspace{-0.8 em}
\begin{thebibliography}{99}
	\bibitem{survey-one}
	D. Lopez-Perez, A. Valcarce, G. de la Roche and J. Zhang, \enquote{OFDMA femtocells: A roadmap on interference avoidance,}  \textit{IEEE Communications Magazine}, vol. 47, no. 9, pp. 41-48, September 2009.
		
	\bibitem{6G-servey}
	G. Gui, M. Liu, F. Tang, N. Kato, and F. Adachi, \enquote{6G: Opening New Horizons for Integration of Comfort, Security and Intelligence,} \textit{IEEE Wireless Communications, Early Access}, 2020.
	
	\bibitem{lwc-2018}
	M. Sinaie, D. Wing Kwan Ng, and E. A. Jorswieck, \enquote{Resource Allocation in NOMA Virtualized Wireless Networks Under Statistical Delay Constraints,}  \textit{IEEE Wireless Communications Letters}, vol. 7, no. 6, pp. 954-957, Dec. 2018.
	
	\bibitem{tvt-2018}
	Q. Wu, W. Chen, D. W. K. Ng and R. Schober, \enquote{Spectral and Energy-Efficient Wireless Powered IoT Networks: NOMA or TDMA?,}  \textit{IEEE Transactions on Vehicular Technology}, vol. 67, no. 7, pp. 6663-6667, July 2018.
	
	\bibitem{jiot-2020}
	X. Liu, H. Ding, and S. Hu, \enquote{Uplink Resource Allocation for NOMA-based Hybrid Spectrum Access in 6G-enabled Cognitive Internet of Things,}  \textit{IEEE Internet of Things Journal, Early Access}, 2020.
	
	\bibitem{tsp-2020}
	L. Salaun, M. Coupechoux, and C. S. Chen, \enquote{Joint Subcarrier and Power Allocation in NOMA: Optimal and Approximate Algorithms,} \textit{IEEE Transactions on Signal Processing}, vol. 68, pp. 2215-2230, 2020.
	
	\bibitem{jiot-2018-noma}
	A. Kiani and N. Ansari, \enquote{Edge Computing Aware NOMA for 5G Networks,}  \textit{IEEE Internet of Things Journal}, vol. 5, no. 2, pp. 1299-1306, April 2018.
	
	\bibitem{lcomm-2018}
	D. Ni, L. Hao, Q. T. Tran and X. Qian, \enquote{Transmit Power Minimization for Downlink Multi-Cell Multi-Carrier NOMA Networks,}  \textit{IEEE Communications Letters}, vol. 22, no. 12, pp. 2459-2462, Dec. 2018.
	
	\bibitem{access-2020}
	Y. Dai and L. Lyu, \enquote{NOMA-Enabled CoMP Clustering and Power Control for Green Internet of Things Networks,} \textit{IEEE Access}, vol. 8, pp. 90109-90117, 2020.
	
	\bibitem{twc-2017}
	Y. Fu, Y. Chen, and C. W. Sung, \enquote{Distributed Power Control for the Downlink of Multi-Cell NOMA Systems,} \textit{IEEE Transactions on Wireless Communications}, vol. 16, no. 9, pp. 6207-6220, Sept. 2017.
	
	\bibitem{fwf-2017}
	S. K. Taskou and M. Rasti, \enquote{Fast Water-Filling Method for Sum-Power Minimization in OFDMA Networks,} \textit{ IEEE Signal Processing Letters}, vol. 24, no. 7, pp. 1058-1062, July 2017.
	
	\bibitem{scell-muser}
	Y. Liu and Y. Dai, \enquote{On the Complexity of Joint Subcarrier and Power Allocation for Multi-User OFDMA Systems,}  \textit{IEEE Transactions on Signal Processing}, vol. 62, no. 3, pp. 583-596, Feb.1, 2014.
	
	\bibitem{wf_2013}
	P. He, L. Zhao, S. Zhou and Z. Niu, \enquote{Water-Filling: A Geometric Approach and its Application to Solve Generalized Radio Resource Allocation Problems,}  \textit{IEEE Transactions on Wireless Communications}, vol. 12, no. 7, pp. 3637-3647, July 2013.
	
	\bibitem{twc-2018}
	T. D. Hoang and L. Bao Le, \enquote{Joint Prioritized Scheduling and Resource Allocation for OFDMA  Wireless Networks,}  \textit{IEEE Transactions on Wireless Communications}, vol. 17, no. 1, pp. 310-323, Jan, 2018.
		
	\bibitem{twc-cran-2018}
	A. Younis, T. X. Tran and D. Pompili, \enquote{Bandwidth and Energy-Aware Resource Allocation for Cloud Radio Access Networks,}  \textit{IEEE Transactions on Wireless Communications}, vol. 17, no. 10, pp. 6487-6500, Oct. 2018.
	
	\bibitem{jiot-2018}
	X. Zhai, X. Guan, C. Zhu, L. Shu and J. Yuan, \enquote{Optimization Algorithms for Multiaccess Green Communications in Internet of Things,}  \textit{IEEE Internet of Things Journal}, vol. 5, no. 3, pp. 1739-1748, June 2018.
	
	\bibitem{tgcn-2019}
	Y. Cheng, J. Zhang, L. Yang, C. Zhu and H. Zhu,  \enquote{Distributed Green Offloading and Power Optimization in Virtualized Small Cell Networks With Mobile Edge Computing,} \textit{IEEE Transactions on Green Communications and Networking}, vol. 4, no. 1, pp. 69-82, March 2020.
		
	\bibitem{tpc}
	G. ~J. ~Foschini, ~ Z. ~Miljanic, \enquote{A simple distributed autonomous power control algorithm and its convergence ,} \emph{IEEE Transaction on Vehicular Technology,} vol. 42, no. 4, 1993.
	
	\bibitem{noma}
	M. Zeng, W. Hao, O. A. Dobre, Z. Ding, and H. V. Poor, \enquote{Power Minimization for Multi-Cell Uplink NOMA With Imperfect SIC,} \textit{IEEE Wireless Communications Letters}, vol. 9, no. 12, pp. 2030-2034, Dec. 2020.
	
	\bibitem{fixed-point}
	V. Pata, \enquote{Fixed-point theorems	and applications}, available online at: https://www.math.washington.edu/acmssem/2011/acms-lecture.pdf
	
	\bibitem{boyd}
	S. Boyd, and L. Vandenberghe \enquote{Convex optimization}, \emph{Cambridge, U.K.:		Cambridge Univ. Press,} 2004.
	
	\bibitem{standards}
	Chi Wan Sung and Kin-Kwong Leung, \enquote{A generalized framework for distributed power control in wireless networks,}  \textit{IEEE Transactions on Information Theory}, vol. 51, no. 7, pp. 2625-2635, July 2005.
	
	\bibitem{pathgain}
	S. Kazemi and M. Rasti, \enquote{Joint power control and sub-channel allocation for co-channel OFDMA femtocells,} \textit{in Proceeding of 2016 IEEE Symposium on Computers and Communication (ISCC), Messina}, 2016, pp. 1171-1176.
	

\end{thebibliography}
\end{document}